\documentstyle[aps,pre,psfig]{revtex}

\def\dt{\delta t}
\def\frot{{\textstyle\frac{1}{2}}}%
\def\dt{\delta t}

\def\meanm#1{\langle #1\rangle}
\begin{document}
\draft
\twocolumn[
\null
\begin{center}
\preprint{HEP/123-qed}
{\large\bf Viscous Flow and Jump Dynamics in Molecular 
Supercooled Liquids:\par I Translations \par }
\vskip 1.5em
Cristiano De Michele${}^{1}$ and Dino Leporini${}^{1,2}$ \cite{byline}\par
{\it
${}^{1}$ Dipartimento di Fisica, Universit\`a di Pisa, V.Buonarroti, 2 
I-56100 Pisa, Italy \par}
{\it
${}^{2}$ Istituto Nazionale di Fisica della Materia, Unit\`a di Pisa \par}
{( Received \today ) \par}
\vskip 0.9em
\begin{minipage}{14.1cm}
\small
The transport and the relaxation properties of a molecular 
supercooled liquid on a isobar is studied by molecular 
dynamics. The molecule is a rigid heteronuclear biatomic system. 
The diffusivity is fitted over four orders of magnitude by the power
law $D \propto \left( T-T_c\right)^{\gamma_{D}}$ with 
$\gamma_{D} = 1.93 \pm 0.02$ and $T_c = 0.458 \pm 0.002$. 
The self-part of the 
intermediate scattering function $F_{s}(k_{max},t)$ exhibits a 
step-like behavior at the lowest temperatures. On cooling, 
the increase of the related relaxation time $\tau_{\alpha}$  
tracks the diffusivity, i.e. $\tau_{\alpha} \propto ( 
k^{2}_{max}D)^{-1}$. At the lowest temperatures fractions of 
highly mobile and trapped molecules are evidenced.
Translational jumps are evidenced. The duration 
of the jumps exhibits a distribution. The distribution of the 
waiting-times before a jump takes place $\psi(t)$ is exponential at 
higher temperatures. At lower temperatures a power-law 
divergence is evidenced at short times, $\psi(t) 
\propto t^{\xi-1}$ with $ 0 < \xi \leq 1$ which is ascribed to 
intermittency. The shear viscosity is fitted by the power law 
$\eta \propto \left(T-T_c\right)^{\gamma_{\eta}}$ with 
$\gamma_{\eta}= -2.20 \pm 0.03$ at the lowest temperatures. At higher 
temperatures the Stokes-Einstein fits the data if {\it stick} boundary 
conditions are assumed. The product $D\eta/T$ increases at lower 
temperatures and the Stokes-Einstein relation breaks down at a 
temperature which is close to the one where the intermittency is 
evidenced by $\psi(t)$. A precursor effect of the breakdown is 
observed which manifests as an apparent stick-slip transition.
\end{minipage}
\end{center}
\par
]

\section{INTRODUCTION}
\label{sec:intro}
The relaxation phenomena and the tranport properties of supercooled 
liquids and glassy materials are topics of current interest  
\cite{noi,review}. It is well known that, on approaching the glass 
transition temperature $T_{g}$ from above, diffusion coefficients and 
relaxation times exhibit remarkable changes of several orders of 
magnitude which are under intense experimental, theoretical and numerical 
investigation. In the high-temperature regime the changes 
 usually track the shear viscosity $\eta$ in the sense that , 
 if $X$ denotes the diffusivity or the inverse of a relaxation time, 
 the product $X \eta/T$ is nearly temperature-independent. In 
 particular, both the Stokes-Einstein, $D \approx kT/6\pi \eta a$, and 
 the Debye-Stokes-Einstein laws, $D_{r} \approx kT/ \eta a^{3}$ are 
 found to work nicely, $D$, $D_{r}$ and $a$ being the translational 
 and the rotational diffusivity and the molecular radius, respectively.
 Differently, in deeply supercooled regimes there is wide evidence that 
the product {\it increases on cooling}  
evidencing the breakdown of the hydrodinamic behavior at molecular 
level and the decoupling by the viscous flow
\cite{rossler,sillescu1,rosslerjpc,rosslerjcp} 
\cite{ediger,tork2,sillescu2,ye,tork3,lepo2,voronel,lepo3,lepo4}. 

Several models suggest that the decoupling between microscopic time 
scales and the viscous flow is a signature of the heterogeneous dynamics 
close to the glass transition, i.e. a spatial distribution of 
transport and relaxation properties  
\cite{sti,opp,douglas,sillescu3}. Interesting alternatives are 
provided by frustrated lattice gas models \cite{antonio} and the 
``energy landscape'' picture 
\cite{review,goldstein,sastry,angell,ruocco}. Most intepretations 
suggest the existence of crossover temperatures below which a change 
of relaxation mechanism must occur \cite{rossler}. These temperatures 
are broadly found around $1.2 T_{g}$, i.e. in the region where the 
critical temperature $T_{c}$ predicted by the mode-coupling theory of 
the glass transition ( MCT ) is found \cite{gotze}. Recent extensions 
of MCT for the shear viscosity to take into account current-fluctuations
in addition to density fluctuations are reported \cite{gotze2}.

During the last years molecular dynamics simulations ( MD ) proved 
to be a powerful tool to investigate supercooled liquids ( for a 
recent review see ref.\cite{kob} ). To date, MD studies investigated 
decoupling phenomena in atomic pure liquids \cite{ruocco} and atomic 
binary mixtures \cite{thiru,barrat,yamamoto1,yamamoto2,harrowell,sharon}.
Most MD studies confirmed that the decoupling is due to dynamic 
heterogeneities
\cite{thiru,yamamoto1,yamamoto2,harrowell,sharon,claudio}.
In fact, "active" \cite{yamamoto2} or "mobile" 
\cite{claudio} regions which largely contribute to set the 
macroscopic average value have been identified. In such regions 
hopping processes, enhancing the transport with respect to the 
hydrodynamic behavior, have been evidenced  \cite{thiru,yamamoto2}. 
The occurrence of jumps in 
glasses has been reported several times in the recent past
\cite{barrat,bernu,wahnstrom,muranaka}. 

We are not aware of MD studies of the decoupling, and, more 
generally, of dynamic heterogeneities in {\it molecular} systems.
This motivated the present and the following paper \cite{demichel2} 
to investigate the transport and the relaxation 
in a three-dimensional, one-component, molecular system. This is 
an important feature since most experimental work is carried out on 
that class of materials. In particular, the decoupling of microscopic 
relaxation from the viscous flow will be addressed in the light of
the increased role played by the hopping processes. The present 
paper is limited to the translational degrees of freedom 
whereas the following paper \cite{demichel2} deals with the 
rotational degrees of freedom.

The paper is organized as follows. In Sec. \ref{sec:model} 
the model and the details of the simulation are presented. The 
results concerning both single-particle and collective properties 
are discussed in Sec. \ref{sec:resul}. The main conclusions are 
summarized in Sec. \ref{sec:concl}.

\section{MODEL AND DETAILS OF SIMULATION}
\label{sec:model}
The system under study is a biatomic molecular liquid. 
The model has been extensively investigated to test the MCT 
predictions \cite{kammererrot1,kammererrot2,kammerertrasl}.
The atoms A 
and B of each molecule have mass $m$ and are spaced  by $d$. 
Atoms on different molecules interact via the 
Lennard-Jones potential:

\begin{equation}
V_{\alpha\beta}(r) = 4\epsilon_{\alpha\beta} \left[ 
{(\sigma_{\alpha\beta}/r)}^{12} 
- {(\sigma_{\alpha\beta}/r)}^6\right],\quad \alpha,\beta \in \{A,B\}
\end{equation}

The potential was cutoff and shifted at $r_{cutoff}=2.5\sigma_{AA}$.
Henceforth,  reduced units will be used. Lenghts are in units of 
$\sigma_{AA}$, energies in units of $\epsilon_{AA}$ and masses in 
units of $m$. The time unit is  $\left(\frac{m 
\sigma_{AA}^2}{\epsilon_{AA}}\right)^{1/2}$
, corresponding to about $2 ps$ for the Argon atom.
The pressure $P$, temperature $T$ and shear viscosity 
$\eta$ are in units of $\epsilon_{AA}/ \sigma_{AA}^{3}$ ,
$\epsilon_{AA} /k_{B}$ and $\sqrt{ m \epsilon_{AA}}/ \sigma_{AA}^{2}$, 
respectively.

The model parameters in reduced units are: 
$ \sigma_{AA}=\sigma_{AB} = 1.0$, $\sigma_{BB}= 0.95$, 
$\epsilon_{AA}= \epsilon_{AB}=1.0$, $\epsilon_{BB}=0.95$, $d = 0.5$, 
$m_A = m_B = m = 1.0$. 
The  $\sigma_{AA}$ and $\sigma_{BB}$ values were chosen 
to avoid crystallization. The sample has 
$N = N_{at}/2 = 1000$ molecules which are accommodated in a 
cubic box with periodic boundary conditions. The viscosity was 
evaluated by using samples of $N=108$ molecules.

We examined the isobar at $P = 1.5$ by the following procedure. 
First, the sample was equilibrated in isothermal-isobaric conditions. 
for a time $t_{eq}$  $t_{eq}$ was at least one order of magnitude 
longer the time needed by the self-part of the intermediate 
scattering function evaluated at the maximum of the static 
structure factor to become smaller than $0.1$.
In this step the equations of motion were integrated by using the 
RATTLE algorithm with Nos\`e-Andersen constant temperature and 
pressure dynamics \cite{allen}. The algorithm is detailed in Appendix  
\ref{app:NPT}. The starting conditions of the equilibration run 
make  the total momentum of the system to vanish and locate the center 
of mass at the center of the  box. The Nos\`e-Andersen Lagrangian 
ensures that the center-of-mass position and the total momentum will 
not change during the run.

The data were collected in a production run in microcanonical 
conditions.  Integration was carried out by the RATTLE algorithm. The 
$\dt$ step ranges from $0.001$ at higher temperatures to $0.004$ at 
lower ones in the production run. The temperatures we investigated 
are $T= 6, 5, 3, 2, 1.4, 1.1, 0.85, 0.77, 0.70, 0.632, 0.588, 
0.549, 0.52, 0.5$. The characterization of the system at $T=0.77$ was 
limited to the diffusivity and the viscosity.
The density at $T=0.5$ was $0.6998$ and decreased 
of a factor of about three at the highest temperatures. To check 
that thermal hystory is negligible, at 
least two independent equilibrations were performed and the 
subsequent production runs compared at the lowest temperatures.

\section{RESULTS AND DISCUSSION}
\label{sec:resul}
The section discusses the results of the study. First, we characterize 
the static properties of the system, i.e. the radial 
distribution function of the center-of-mass $g(r)$ and the static 
structure factor $S(k)$. Then, the single-particle and the collective 
dynamical  properties of the system  will be presented.

\subsection{Static properties}
The radial distribution function  of the center of mass $g(r)$ is 
 defined as:

\begin{equation}
g(r) = \frac{1}{4 \pi N \rho r^2}
\sum_{i\neq j} \langle\delta(|{\bf R}_{i}(0) - {\bf 
R}_{j}(0) - {\bf r}|)\rangle
\end{equation}

\begin{figure}
\psfig{file=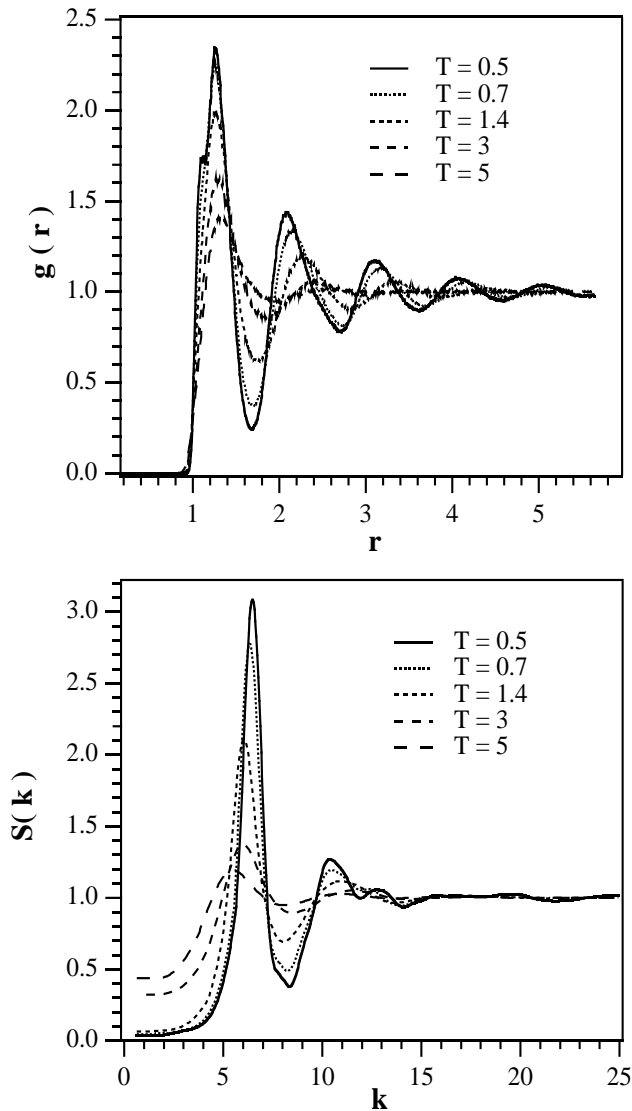,width=9cm,height=15cm}
\caption{Radial distribution function $g(r)$ ( top ) and static 
structure factor $S(k)$ ( bottom ) of the molecular 
center of mass for selected temperatures.}
\label{grskFig}
\end{figure}

\noindent $\rho$ is the average density and ${\bf R}_{i}(t)$ is the 
position of the center of mass of the $i-$th molecule at time $t$.
 Representative plots of  $g(r)$ at different 
temperatures are shown in Fig. \ref{grskFig} . The pattern is 
typical of a disordered system. At $T = 0.5$,  $g(r)$ exhibits 
the maximum  at $r \cong 1.2$ and the minimum at $r \cong 1.6$. 
On increasing the temperature, the peaks broaden and shift at 
higher $r$ values. The shoulder which is observed at $r \cong 1.1$ 
at the lowest temperatures has been already observed \cite{private} 
and may be ascribed to local ``T-shaped'' or cross configurations 
of the molecules \cite{hansen}. 

The static structure factor $S(k)$ is defined as 

\begin{equation}
S( k) =  \frac{1}{N} \langle \rho_{k} \rho_{-k} \rangle
\end{equation}

\noindent $\rho_{k}$ is the Fourier transform of the density.
Representative plots of $S(k)$ at different 
temperatures are shown in fig. \ref{grskFig}. 

The above results rule out possible crystallizations of the 
sample. This is anticipated by the absence of anomalies 
in the temperature dependence of the density.
Direct inspection of several snapshots of the 
molecular configurations also supported the conclusion and 
did not reveal any orientational ( liquid-crystalline ) order. 
The absence of orientational order was also evidenced by the 
long-time behaviour of the rotational correlation function 
\cite{demichel2}. 

\subsection{Single-particle dynamics}

The section will discuss the single-particle dynamics. 
At short time scales one quantity of interest is the 
velocity self-correlation function of the center of mass :

\begin{equation}
C_{vv}(t) = \frac{1}{N}
	\sum_{i=1}^N\langle {{\bf v}_i(t)}\cdot{{\bf v}_i(0)}\rangle
\end{equation}

\begin{figure}
\psfig{file=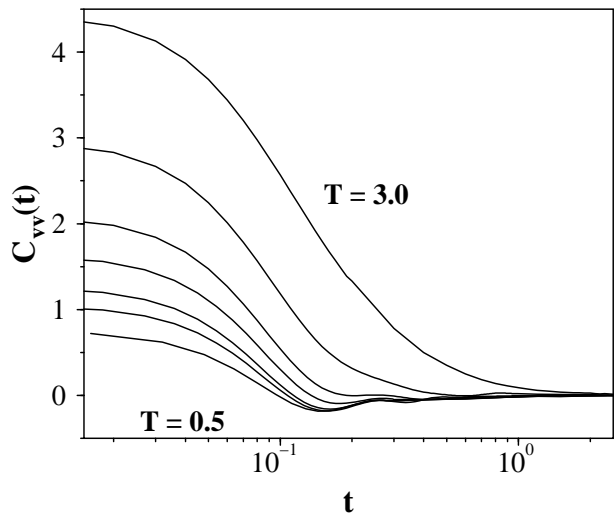,width=8cm,height=7cm}
\caption{Velocity self-correlation function of the center of 
mass. The curves refer to $ T = 3, 2, 1.4, 1.1, 
0.85, 0.7, 0.5$.}
\label{Cvv}
\end{figure}

\noindent $\meanm{A(t)}$ denotes a proper time average of $A(t)$.  
Fig. \ref{Cvv} plots $C_{vv}$ for different temperatures.

At very short times $C_{vv}(t) \cong <v^2>(1-<\Omega^2>t^2/2 +\ldots )$,
where $<\Omega^2>$ is the Einstein frequency \cite{hansen}.
At the lower temperatures the damped oscillatory motion of the molecule 
inside the cage where it is accommodated becomes apparent.
The effect signals that on a short time scale the supercooled liquid
behaves as a solid. 

The mean squared displacement of center of mass, $R(t)$ provides a 
first view of the translational motion of the molecules at 
intermediate and long time scales. It is defined as:

\begin{equation}
R(t) = \frac{1}{N} \sum_{i=1}^{N} \meanm{{\vert {\bf R}_i(t) - {\bf 
R}_i(0)\vert}^2}.
\label{rms}
\end{equation}

\begin{figure}
\psfig{file=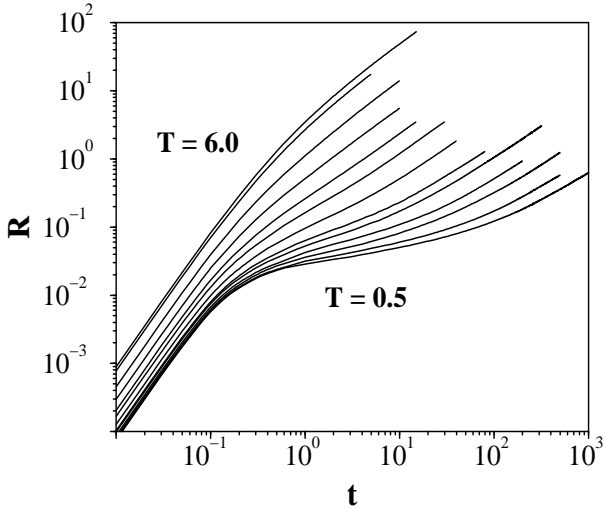,width=8cm,height=7cm}
\caption{Mean squared displacement of molecular center-of- 
mass $R(t)$ for all the temperatures under investigation but $T=0.77$.}
\label{drSq}
\end{figure}

\noindent Plots of $R(t)$ at different temperatures are shown 
in fig.\ref{drSq}. At short times the motion is ballistic and 
$R(t) \propto  t^2$.  At intermediate times $R(t)$ exhibits a 
crossover regime which is strongly temperature dependent.  
At higher temperature it reduces to a knee joining the 
ballistic and the diffusive regimes. At lower temperatures  
the molecule is effectively trapped inside the cage of the first 
neighbours and $R(t)$ exhibits a plateau. At long times the motion 
is diffusive and  $R(t) = 6 D t$ where $D$ is the diffusion constant.

\begin{figure}
\psfig{file=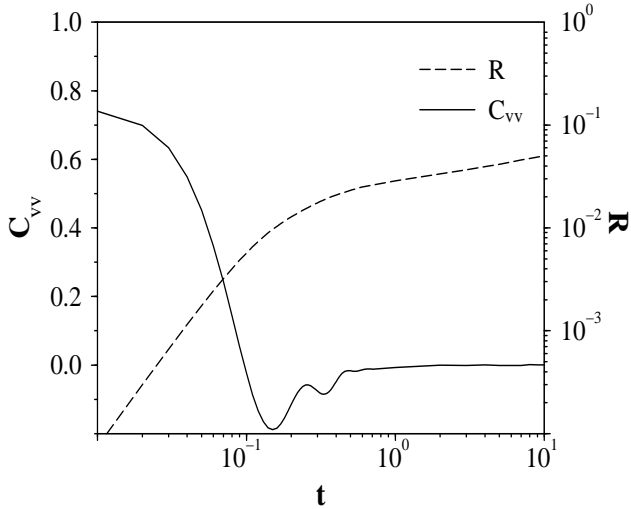,width=8cm,height=7cm}
\caption{Comparison of the mean squared displacement $R(t)$ and 
the velocity correlation function $C_{vv}(t)$ at $T=0.5$. 
The onset of the damped oscillations takes place at the
beginning of the plateau of $R(t)$: the molecule  is colliding 
with the cage of the first neighbours.}
\label{CvvdrSq}
\end{figure}

The occurrence of oscillatory motion during trapping 
is shown explicitely in fig.\ref{CvvdrSq} where the mean 
squared displacement and the velocity correlation function 
are compared at $T=0.5$. The plot evidences that velocity 
correlations got lost within the lifetime of the cage.

The translational diffusion coefficient $D$ is evaluated 
by the Einstein relation \cite{allen,hansen}:

\begin{equation}
D = \lim_{t\to\infty} \frac{R(t)}{6t}
\end{equation}

\begin{figure}
\psfig{file=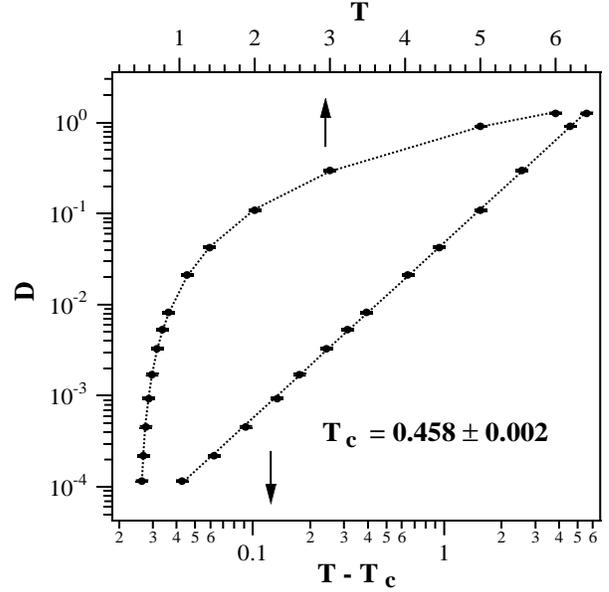,width=8cm,height=8cm}
\caption{Temperature dependence of the translational diffusion 
coefficient $D$. The dashed line is a fit with the power law 
eq.\ref{powerlaw} with $\gamma_{D} = 1.93 \pm 0.02$ , 
$T_c = 0.458 \pm 0.002$ and $C_{D} = 0.0481 \pm 0.0004$.}
\label{Dtrasl}
\end{figure}

\noindent In Fig. \ref{Dtrasl} the temperature dependence of $D$
is shown. The plot emphasizes the scaling property of $D$ which 
is nicely described over four decades by the power law :

\begin{equation}
D = C_{D}\left( T-T_c\right)^{\gamma_{D}}
\label{powerlaw}
\end{equation}

\noindent Theoretical justification of eq.\ref{powerlaw} is provided 
by MCT \cite{gotze}. In particular, the ideal MCT  predicts 
the inequality $\gamma_{D} > 1.5$. Our best fit 
values are $\gamma_{D} = 1.93 \pm 0.02$ and 
$T_c = 0.458 \pm 0.002$, $C_{D} = 0.0481 \pm 0.0004$. For the 
same biatomic system with $N=500$ and $P=1$ it was found 
$\gamma_{D} = 2.2$ and $T_c = 0.475$ \cite{kammererrot1}.

It must be noted that at lower temperatures the above 
power law is expected to underestimate the diffusion coefficient 
\cite{kammerertrasl}. The larger diffusivity with respect to the ideal 
MCT prediction close to $T_{c}$ is not 
surprising. According to this approach, density-density 
correlations do not vanish at long times below $T_{c}$ leading to a 
divergence of $D$ at $T_{c}$. 
The extended MCT points out that hopping processes  provide 
an effective mechanism to relax the density fluctuations at long time 
and make the diffusivity finite close to $T_{c}$. This point will be 
explicitely discussed later.

To characterize further the single-particle dynamics we 
evaluated the self-part of the intermediate scattering 
function :

\begin{equation}
F_s(k,t) = \langle exp[i{\bf k}\cdot 
({\bf R}_{j}(t) -{\bf R}_{j}(0) )] \rangle 
\label{intermediate}
\end{equation}

\begin{figure}
\psfig{file=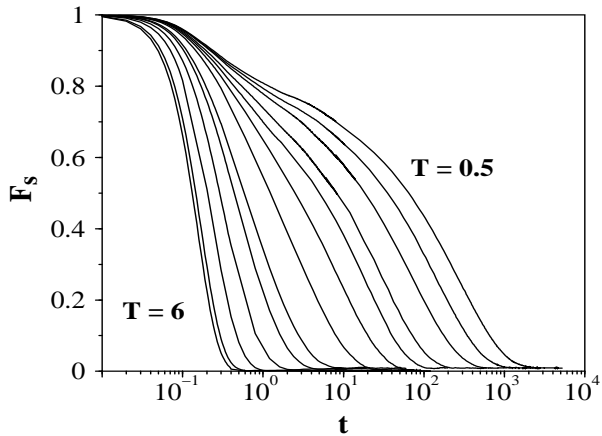,width=8cm,height=6cm}
\caption{Self-part of the intermediate scattering function 
$F_s(k_{max},t)$. The curves refer to all the temperatures under
investigation but $T=0.77$.}
\label{FsFig}
\end{figure}

In fig. \ref{FsFig} $F_s(k,t)$ is plotted for all the temperatures we 
investigated at $k=k_{max}$. $k_{max}$ is the position of the main 
peak of the static structure factor $S(k)$ at temperature $T$. We note 
that $F_s(k,t)$ always decays to zero and is independent of the 
thermal hystory of the sample ( i.e. independent runs lead to the 
same result ) signaling the equilibration of 
 the sample. At higher temperatures the decay is virtually
exponential at long times, whereas at lower temperatures a 
two-step decay is observed.

The plateau which is observed at lower temperatures in 
$F_s(k_{max},t)$ is due to the trapping of the molecules. This is 
seen by comparing fig.\ref{drSq} with fig.\ref{FsFig}. Precise 
predictions on the scaling features of the plateau of 
$F_s(k_{max},t)$ are made by MCT \cite{gotze}. A thorough 
comparison for the present model is found in ref.\cite{kammerertrasl}.

\begin{figure}
\psfig{file=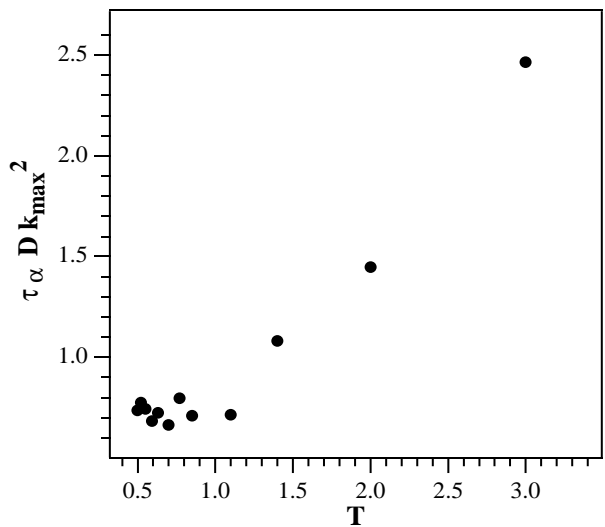,width=8cm,height=7cm}
\caption{Temperature dependence of the product 
$\tau_{\alpha} Dk^{2}_{max}$.}
\label{taualfa}
\end{figure}

At longer times the cage where the molecule is trapped opens and the 
density-density correlations vanish. The long-time decay of 
$F_s(k_{max},t)$  is well fitted by the stretched exponential 
$ A exp(-t/\tau)^{\beta}$. Stretching is appreciable at lower 
temperatures ( $\beta_{0.5} = 0.677$ ) and becomes negligible 
( $\beta \approx 1$ ) for $T > 0.85$. 

The knowledge of $F_s(k_{max},t)$ offers the 
opportunity to investigate the time scale of structural 
relaxation being usually denoted by the $\alpha$-relaxation time 
$\tau_{\alpha}$. We have evaluated  it by the 
condition $F_s(k_{max},\tau_{\alpha}) = 1/e \cong 0.3679$. At low 
temperatures a nice fit is provided by $\tau_{\alpha} = 
C_{\alpha}(T-T_{c})^{\gamma_{\alpha}}$, being $T_{c}$ the same 
best-fit value drawn by the diffusivity.
The best fit value of the exponent is $\gamma_{\alpha} =
 -1.89 \pm 0.05$ which is quite close to $\gamma_{D} = 1.93$. 
 Diffusive motion in simple 
liquids leads to the exponential decay of $F_s(k,t)$  
with time costant $\tau = 1/Dk^{2}$ \cite{hansen}. 
Even the decay at long times is better described 
by a stretched exponential, the approximate equality $\gamma_{\alpha} 
 \cong \gamma_{D}$ prompted us to investigate the 
temperature dependence of the product $\tau_{\alpha} Dk^{2}_{max}$. 
The results are shown in fig.\ref{taualfa}. For $T<1.1$ the product 
approaches the constant value $0.72 \pm 0.05$. In this range 
$\tau_{\alpha}$ and $D$ change of more than two orders of magnitude.
Alternative definitions of 
$\tau_{\alpha}$, e.g. involving the area below $F_{s}(k_{max},t)$ 
do not affect significantly the result.
Removing the $k_{max}^{2}$ term results in sligthly
poorer fit. The $k^{2}$ scaling of the primary relaxation time 
$\tau^{s}_{k}$, evaluated as the area below $F_{s}(k,t)$, 
was already noted in the hard sphere system at the critical 
packing fraction in the region $1.5 \leq  ka \leq 30$ 
\cite{fuchs} . In particular, it was found
$\tau^{s}_{k} Dk^{2} \sim 1.08$ in good agreement with 
our result and $F_{s}(k,t)$ exhibits stretching 
with $\beta(k_{max}) \sim 0.8$ .

The above findings agree with some predictions of MCT: 
i) $T_{c}$ must be independent of the quantity under 
study, ii) $\gamma_{\alpha}$ must be equal to $\gamma_{D}$. 
The equality $\gamma_{\alpha} = \gamma_{D}$ must be taken with great
caution. Other studies on the same model system with $N = 500$ and 
$P=1$  found $\gamma_{\alpha} > 
\gamma_{D}$ even if fitting the diffusivity and 
$\tau_{\alpha}$ provide the same $T_{c}$ value \cite{kammerertrasl}.

\begin{figure}
\psfig{file=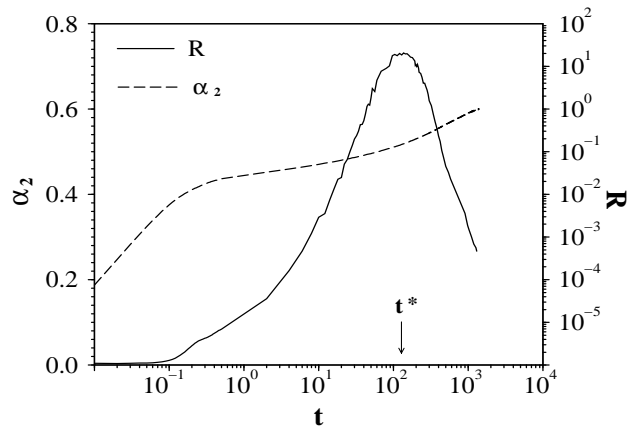,width=8cm,height=6cm}
\caption{Comparison of the non-gaussian parameter $\alpha_2$  and 
the mean squared displacement $R(t)$ at $T=0.5$. The maximum of the 
non-gaussian parameter occurs at $t^{*}$. $t^{*}$ provides an 
estimate of the trapping time and is comparable with $\tau_{\alpha}$.}
\label{nongaussdrSqFig}
\end{figure}

Additional information on the single-particle dynamics
is provided by the self part of the Van Hove function 
$G_{s}({\bf r},t)$ \cite{hansen}:

\begin{equation}
G_{s}(r,t) = \frac{1}{N}\meanm{\sum_{i=1}^N 
\delta({\bf r}+{\bf R}_i(0)-{\bf R}_i(t))}
\label{vanhove}
\end{equation}

\noindent In isotropic liquids the Van Hove function depends only on 
the modulus $r$ of ${\bf r}$. The interpretation of $G_{s}(r,t)$ is 
direct. The product $G_{s}({\bf r},t) \cdot 4\pi r^{2}$ is the 
probability that the molecule is at a distance between $r$ and 
$r+dr$ from  the initial position after a time t.

\begin{figure}
\psfig{file=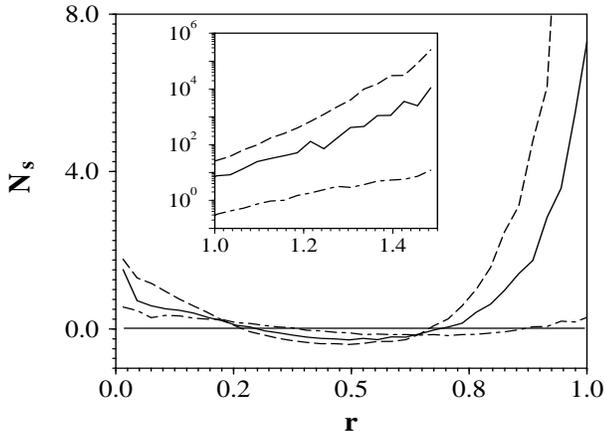,width=8cm,height=6cm}
\caption{ The ratio $N_{s}(r) = (G_s(r,t^*) - 
G_s^g(r,t^*))\break/G_s^g(r,t^*)$ 
plotted at $T=0.5$(dashed line), $T=0.588$ (solid line), $T=0.85$ 
(dot-dashed line). Inset: same quantity on a logarithm scale.}
\label{NFig}
\end{figure}

The shape of $G_{s}(r,t)$ in the system under study does not reveal 
particular features \cite{kammerertrasl}. In particular, differently 
from other studies \cite{barrat}, no secondary peak at  $r \approx 1$, 
the nearest-neighbor distance, is found to 
support the conclusion that hopping is occurring. 
A deeper insight is gained by comparing $G_{s}(r,t)$ 
with the gaussian approximation:

\begin{equation}
G^g_s(r,t) = [3/2\pi\meanm{r^2(t)}]^{3/2}\exp(-3r^2/2\meanm{r^2(t)})
\label{gaussGsEq}
\end{equation}

\noindent Eq. \ref{gaussGsEq} is the correct limit of $G_{s}(r,t)$ at 
short  ( ballistic regime, $\meanm{r^2(t)} = 3kT/m t^{2}$ ) 
and long times (  diffusion regime, $\meanm{r^2(t)} = 6 D t$ ) 
\cite{hansen}.  Thirumalai and Mountain \cite{thiru} and more 
recently Kob et al. \cite{claudio} noted  discrepancies between 
$G_{s}(r,t)$  and $G^g_s(r,t)$ in supercooled atomic mixtures. 
It is found that $G_{s}(r,t)$  exceeds $G^g_s(r,t)$ at short 
and long $r$ values. The effect is particularly evident at $t \approx t^*$
where $t^{*}$ is the time where the non-gaussian parameter 

\begin{equation}
\alpha_2(t) = 3\meanm{r^4(t)}/5\meanm{r^2(t)}^2 - 1,
\end{equation}

\noindent reaches the maximum value \cite{kammerertrasl}. 
At $t^{*}$ the stochastic properties of $\bf{r}$  
differ by the ones of a gaussian variable to the maximum.
A plot of $\alpha_2(t)$ is shown in fig.\ref{nongaussdrSqFig} 
where it is compared to the mean squared displacement $R(t)$ 
at $T=0.5$. The maximum of $\alpha_2(t)$ at $t =t^{*}$ is located
between the trapping and diffusive regimes.
The decrease of $\alpha_2(t)$ for $t< t^{*}$ is due to the recovery 
of the gaussian form of $G_{s}(r,t)$ in the trapping regime
where molecules undergo a nearly oscillatory motion in the 
cages where they are accommodated ( fig. \ref{Cvv} ).
$t^{*}$ provides an estimate of the trapping 
time and in fact is comparable to the $\alpha$-relaxation time.
For $t \approx 0.2-0.3$, corresponding to the second 
maximum of the velocity correlation function ( fig.\ref{CvvdrSq} ), a 
small step is observed. The same feature of the non-gaussian 
parameter was observed in a MD work on a dense bidimensional liquid 
\cite{harrowell2}.

The deviations of the Van Hove function by the gaussian limit at short 
and long $r$ values which were observed for $t \approx t^*$ shows 
that the sample has fractions of both trapped and highly mobile atoms
\cite{thiru,claudio}. It has been suggested that the dynamics of the 
latter is conveniently described by hopping processes \cite{thiru}. 
To study the possible deviations of the Van Hove function in the 
present {\it molecular} system, we considered the ratio \cite{claudio}:

\begin{equation}
N_{s}(r) = \frac{G_s(r,t^*) - G_s^g(r,t^*)}{G_s(r,t^*)}
\label{normalratio}
\end{equation}

\noindent Fig. \ref{NFig} shows the ratio $N_{s}(r)$ for different 
temperatures. 
It exhibits increasing positive deviations at both short and large 
$r$ values on cooling. This supports the conclusion that the dynamical 
heterogeneities evidenced in atomic two-phase systems are also present in 
{\it molecular} one-phase systems \cite{thiru,claudio}. 

Non-vanishing $N_{s}(r)$ values could be anticipated
by noting that the self-parts of the intermediate scattering 
function and the Van Hove function are related to each other by:

\begin{equation}
F_s(k,t) = \int_0^{+\infty} G_s(r,t)\, 4\pi r^2\frac{\sin kr}{kr} dr
\label{intevsVH}
\end{equation}

\noindent According to eq.\ref{intevsVH}, the main contributions to 
$F_s(k_{max},t)$ come roughly from  $r < r^{*}$ with $r^{*} 
\approx 2\pi/k_{max} \approx 1$.
If the discrepancies between $G_s(r,t^*)$ and $G_s^g(r,t^*)$ for $r<1$ 
were small, $F_s(k_{max},t) \approx exp(-Dk_{max}^2 t)$ for 
$t \approx t^{*}$. Instead, $F_s(k_{max},t)$
is found to decay as a stretched exponential ( see fig.\ref{FsFig} 
and related discussion ). 

\begin{figure}
\psfig{file=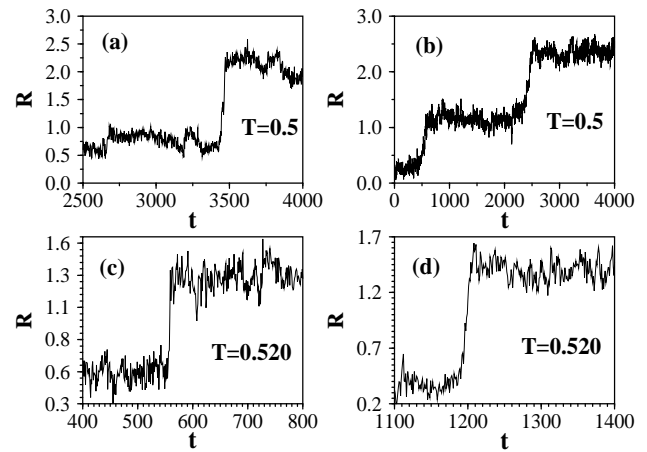,width=8cm,height=6cm}
\caption{Squared displacements $R$ of selected molecules at 
$T=0.5$ and $T=0.520$. Note the distribution of the time needed 
to complete the jumps.}
\label{traslJumpFig}
\end{figure}

Since the positive tail of $N(r)$ at large 
$r$ starts at $r \approx 1 $ i.e. about the molecular size, it 
is tempting to ascribe it to jump motion. Indeed, by inspecting the 
particle trajectories jumps are found. Examples are shown in fig. 
\ref{traslJumpFig}. It was noted that the jump duration exhibits a
distribution ( roughly between $20-500$ time units ), suggesting that 
different degrees of cooperativity are involved \cite{bernu}. 
The translational jumps which are detected are relatively slower than 
the rotational jumps (flips of about $180^{\circ}$  ) which 
take about $6-7$ time units with little or no distribution 
\cite{demichel2,kammererrot1}. Furthermore, translational 
jumps are found to be much rarer than rotational ones
\cite{demichel2}. In fact, differently from 
the translational case, the large amount of molecule
flips manifests in a secondary peak of the proper Van Hove function
\cite{demichel2,kammererrot1,kammerertrasl}. 

The above remarks point to the conclusion that translational jumps are not 
obviously related to rotational ones.  Their different character will be 
more clearly evidenced by studying and comparing the distributions of
the waiting-times, i.e. the lapse of time between successive jumps 
( see below and ref. \cite{demichel2}).

\begin{figure}
\psfig{file=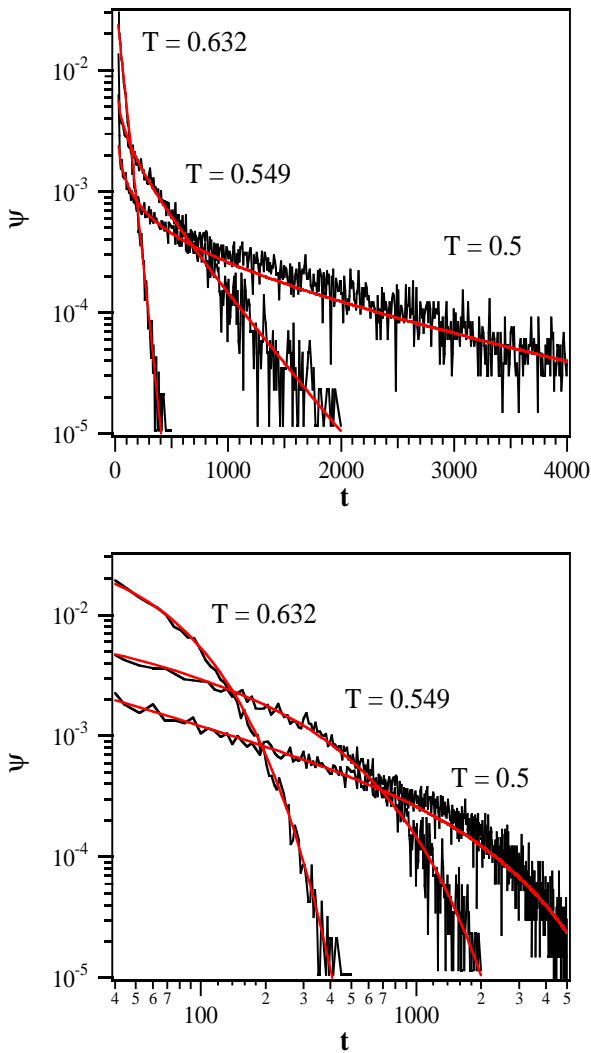,width=8cm,height=14cm}
\caption{Long-time (top ) and short-time ( bottom ) behavior of the 
waiting-time distribution $\psi(t)$ at different temperatures. 
The solid lines superimposed to the curves is eq.\ref{psi} 
with $\xi = 0.49$, $\tau = 2550$ ( $T = 0.5$ ),
$\xi = 0.63$, $\tau = 420$ ( $T = 0.549$ ) 
and $\xi = 1$, $\tau = 49$ ( $T = 0.632$ ).}
\label{wtdfig}
\end{figure}

The  evaluation of the waiting-time distribution $\psi(t)$ relies on 
the explicit definition of a jump event. In the present study a 
molecule jumps at time $t$ if the displacement between $t$ and 
$t+\Delta t ^{*}$ ($\Delta t ^{*} = 24$) exceeds $\sqrt{\Delta R^{*}} 
= \sigma_{AA}/2 = 0.5$. To avoid multiple counting, the molecule 
which jumped at time $t$ is forgotten up to time $t+\Delta t ^{*}$.
Possible spurious countings due to fast rattling motion are minimized 
by averaging each displacement with the previous and the next one.
These are typically spaced  by $6-8$ time units, depending on the 
temperature \cite{muranaka}. It must be noted that after a time 
$\Delta t ^{*}$ the average diffusive displacement
is smaller than $\Delta R^{*}$, e.g. $\Delta R^{diff} = 6D\Delta t ^{*}
 = 0.017$ at $T=0.5$ and $\Delta R^{diff} = 0.032$ at $T=0.520$.
We validated the jump search procedure by 
inspecting several single-molecule trajectories. In particular, it was 
checked that molecular vibrations in local cages
are not misinterpreted as jumps contributing to 
$\psi(t)$ at short-times. Some attempts 
to investigate how $\psi(t)$ depends on the above definition 
were made. Since they require wide statistics, the issue will not be 
touched in the present paper. It will be discussed in a more 
detailed way elsewhere. 

\begin{figure}
\psfig{file=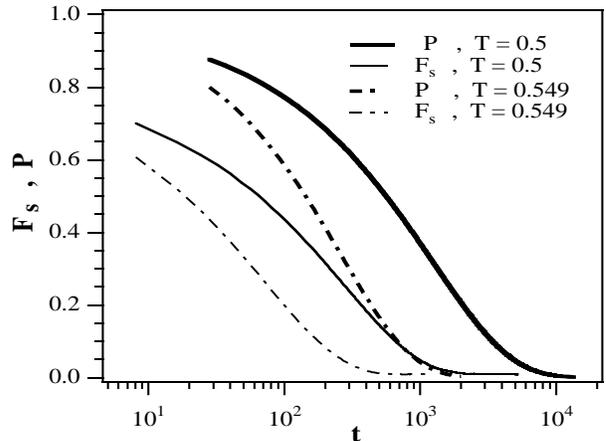,width=8cm,height=6cm}
\caption{Comparison between the probability that 
the waiting time between consecutive jumps is larger than $t$,
$P(t)$, and  $F_{s}(k_{max},t)$ at $T=0.5$ and $0.549$.}
\label{PFs}
\end{figure}

Fig.\ref{wtdfig} shows the waiting-time distribution $\psi(t)$ at 
different temperatures. At high temperature $\psi(t)$ is exponential.
On cooling, the exponential decay is replaced at short times by
a slowly-decaying regime. We fitted the overall decay by the function:
 
 \begin{equation}
\psi(t) =  \left[\Gamma (\xi )\tau^{\xi}\right]^{-1}
t^{\xi-1} e^{ -t / \tau } \hspace{1cm} 0 < \xi \le 1
\label{psi}
\end{equation}
 
\noindent The choice is motivated by the remark that 
in glassy systems rearrangements are rare events due to the 
constraints hampering the structural relaxation. It is believed that
intermittent behaviour in particle motion develops on cooling 
\cite{sharon,sjogren,hubbard,odagaki}. A signature of 
intermittence is the power law decay of $\psi(t)$ ( for 2D liquids 
see \cite{muranaka} ) and related quantities such as the first-passage 
time distribution \cite{sharon}. The exponent $\xi$ of eq.\ref{psi}
has a simple interpretation. If a dot on the time axis 
marks each relaxation event ( a jump ), the fractal dimension 
of the set of dots is $\xi$. For $\xi < 1$, it is found
$\psi(t) \propto t^{\xi-1}$ at short times \cite{sjogren,hilfer}. 
Modelling the long-time decay of $\psi(t)$ is less obvious. As a first 
guess, if the distribution of events becomes nearly uniform ( $\xi =1$) 
beyond a time scale $\tau$ and homogeneous across the sample, $\psi(t)$ 
recovers the exponential form.

The best fits at $T=0.5,0.549$ and $0.632$ are shown in 
fig.\ref{wtdfig}. The increase of temperature results in a weak 
increase of the exponent $\xi$ and a more marked decrease of $\tau$.
 It was found that eq.\ref{psi} fits $\psi(t)$ also by setting 
 $\sqrt{\Delta R^{*}}$ and $\Delta t ^{*}$ in the ranges $0.4-0.7$ and  
$24-48$, respectively. This suggests that the character of the decay of 
$\psi(t)$ is not strongly dependent on how a jump is defined. 
Nonetheless, the best fit values of the parameters $\xi$ and $\tau$ 
depend on $\sqrt{\Delta R^{*}}$ and $\Delta t ^{*}$ differently. 
For example, at $T=0.5$,  if $\sqrt{\Delta R^{*}}$ 
changes from $0.5$ to $0.6$ with $\Delta t ^{*} = 24$, 
$\xi$ does not change within the errors whereas $\tau$ increases of 
a factor of about $2.4$. $\tau$ also increases by decreasing the time 
$\Delta t ^{*}$ allowed to perform the displacement 
$\sqrt{\Delta R^{*}}$.
The increase of $\tau$ is understood by noting that increasing the 
threshold $\sqrt{\Delta R^{*}}$ or, alternatively decreasing 
$\Delta t ^{*}$, reduces the number of sudden displacements which 
comply with the definition of "jump" and then increases the waiting time
before a new jump occurs. Interestingly, at $T=0.5$ $\psi(t)$ exhibits
small but reproducible deviations from eq.\ref{psi}. They suggest that
the long-time decay is {\it faster } than the exponential one. If the 
exponential decay is replaced by a gaussian one, the fit improves quite
a lot and the $\xi$ exponent changes from $0.49$ to $0.45$. Even if 
this refinement is rather suggestive, we prefer to consider it as ad-hoc
at the present level of understanding.

\begin{figure}
\psfig{file=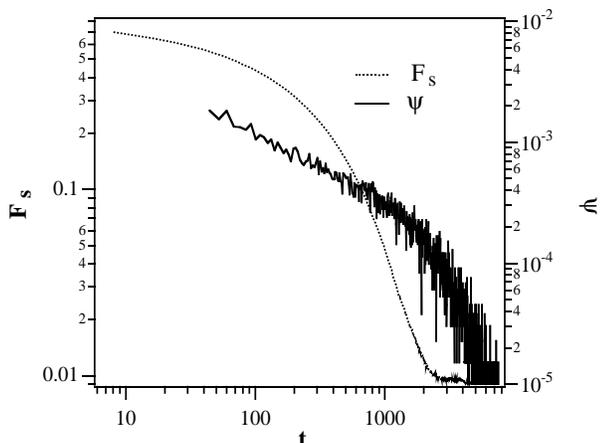,width=8cm,height=6cm}
\caption{Comparison between $\psi(t)$ and  $F_{s}(k_{max},t)$ 
at $T=0.5$.}
\label{wtdFs}
\end{figure}

Further insight on $\psi(t)$ is offered by the study of the 
interplay between density fluctuations and the occurrence of jumps.
This is of interest since, according to the extended version of 
MCT the dynamic transition 
occurring at $T_{c}$ from an ergodic to a non-ergodic state 
is actually smeared by hopping processes \cite{gotze}. 
Fig.\ref{PFs} compares at low temperatures the long-time tail of
$F_{s}(k_{max},t)$ and the probability that no jumps occurred before 
$t$, $P(t) = \int_{t}^{\infty}\psi(x) \,dx$. It is seen that the
density self-correlations vanish at times when $P(t)$ is still 
meaningful ( when $F_{s}(k_{max},t) \sim 0.1$ $P(t) \sim 0.5$ at $T=0.5$). 
There is also some evidence that the fraction of waiting times longer 
than the time needed to make $F_{s}(k_{max},t)$ vanishing ( e.g. 
$F_{s}(k_{max},t) < 0.1$ ) increases on cooling.
Strong analogies with the above findings are found in a recent MD work 
on the viscous silica melt \cite{silica}. It was reported that the 
probability that a bond between a silicon and an oxygen atom 
which was present at time zero is still present at time $t$ vanishes 
much later than $F_{s}$ even at high temperature. 

Then, at lower temperatures two restructuring regimes are identified 
depending on the length of the waiting time with respect to 
$\tau_{\alpha}$.  This is shown in fig. \ref{wtdFs}. For 
waiting times shorter than $\tau_{\alpha}$ molecules jump in a
nearly frozen environment. Escape from the cage becomes more 
difficult by lowering $T$ and, expectedly, with a distribution of 
rates leading to the non-exponential decay of $\psi(t)$.
For waiting times longer than $\tau_{\alpha}$ a larger restructuring of
the surrounding environment takes place. This averages the dynamical 
heterogeneities and leads to the long-time decay of $\psi(t)$.

It is tempting to note that in the time window where $\psi(t)$ 
exhibits the power law decay MCT predicts that 
$F_{s}(k,t)$ itself decays as the von Schweidler power law 
\cite{gotze}:

\begin{equation}
F_{s}(k,t) = f^c_{s}(k) - h^s(k) ( \frac{t}{\tau} )^b + \ldots
\label{vonschweidler}
\end{equation}

\noindent where $f^c_{s}(k),h^s(k)$ and $b$ are constants. 
$F_{s}(k_{max},t)$ must be compared to $P(t) \approx A - B t^{\xi} 
+ \ldots$ in the region of interest. $A$ and $B$ are constants.
If one estimates $b$ via the $\beta$ parameter of the stretched exponential
fit of the long-time decay of $F_{s}(k_{max},t)$, it is found $b \cong 
\beta = 0.68$ at $T=0.5$. This must be compared to $\xi=0.45$.
To make clearer to what extent the two time fractals are related to 
each other the analysis should be refined \cite{kammerertrasl}.
This is beyond the purposes of the present paper. However, we note 
that for times shorter than about $\Delta t ^{*} = 24$ $\psi(t)$ and 
then $P(t)$ cannot be defined since most jumps are not completed.

\subsection{Collective dynamics: the shear viscosity}

The shear viscosity $\eta$ was evaluated by using the Einstein 
relation \cite{allen1}:

\begin{equation}
\eta = \frac{1}{2VkT} \lim_{t\to+\infty} \frac{1}{t}
\meanm{\Delta A(t)^2}
\label{einstein1}
\end{equation}

\noindent where:

\begin{equation}
\Delta A(t)  = V \int_0^{t} {\cal P}_{\alpha\beta}(t') dt' 
\label{einstein3}
\end{equation}

\noindent ${\cal P}_{\alpha \beta}$ is one off-diagonal component 
of the pressure tensor \cite{hansen,allen1,note} ( in practice 
$\Delta A(t)$ is the average over the three possible choices 
$\alpha \beta = xy, xz, yz$ ). 

If the quantity ${\cal P}_{\alpha\beta} V$ is 
evaluated based on the motion of individual atoms comprising the 
molecules in the system we have \cite{allen2}
 
\begin{equation}
{\cal P}_{\alpha \beta}^{at}(t) V = 
\sum_{i=1}^{N}  m_{i} v_{i \alpha}v_{i \beta} + 
\sum_{i=1}^{N} \sum_{j > i}^{N} f_{\alpha ij} (r_{i \beta} - r_{j \beta})
\label{PtensatomDef}
\end{equation}

\noindent The sums involve components ( denoted by greek letters ) of
${\bf v}_{i}$, ${\bf r}_{i} $ and ${\bf f}_{ ij}$  which are 
the velocity and the position of the $i$-th atom having mass $m_{i}$ 
and the force between the atoms $i$ and $j$ ( assumed pairwise 
additive ), respectively. Eq.\ref{PtensatomDef} is not affected by the 
periodic boundary conditions employed in MD simulations. 

An interesting alternative to the above 
atomic representation of the pressure tensor is the molecular one 
which replaces ${\cal P}_{\alpha\beta}^{at}$ with the symmetric part 
of the tensor \cite{allen2}

 \begin{equation}
{\cal P}_{\alpha\beta}^{mol}(t) V = \sum_{i=1}^{N}  M_{i} V_{i 
\alpha}V_{i \beta} + 
\sum_{i=1}^{N} \sum_{j > i}^{N} F_{\alpha ij} (R_{i \beta} - R_{j \beta} )
\label{PtensmolDef}
\end{equation}

\noindent The sums now involve components of
${\bf V}_{i}$, ${\bf R}_{i} $ and ${\bf F}_{ ij}$  which are 
the centre-of-mass velocity and the centre-of-mass position of 
the $i$-th molecule having mass $M_{i}$ 
and the total force between the molecules $i$ and $j$, respectively.
Evaluating the atomic pressure tensor via eq.\ref{PtensatomDef} is a 
little more efficient than the alternative one and was adopted in the 
present study. Nonetheless, we found that the 
two representations exhibit the same convergence when evaluating 
eq.\ref{einstein1} and yield identical results in agreement with 
previous studies \cite{cui}.

\begin{figure}
\psfig{file=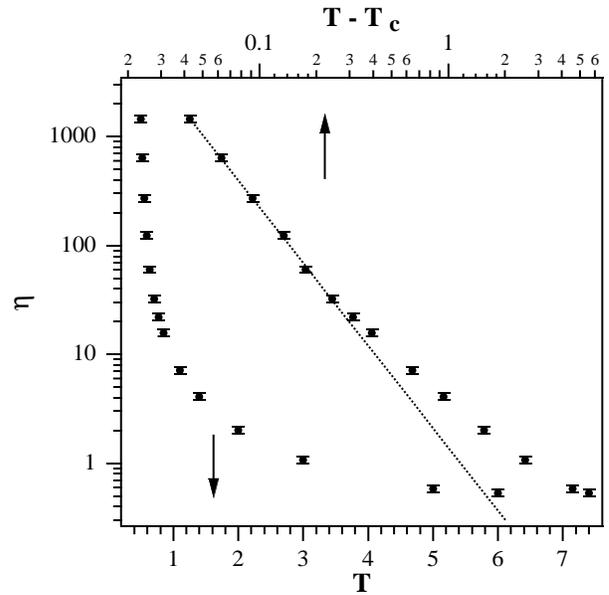,width=8cm,height=8cm}
\caption{Temperature dependence of the shear viscosity. The 
superimposed line has slope $\gamma_{\eta}= -2.20 \pm 0.03$.}
\label{viscoFig}
\end{figure}

The evaluation of the shear viscosity is particularly time-consuming 
being a collective property which involves many-particle correlations.
In fact, if one evaluates a generic $N$-particle property by a run 
spanning a time interval as long as $t$, the relative error $r$ is 
of the order of $(\tau_{c}/t)^{1/2}$, $\tau_{c}$ being 
the correlation time. The error is quite larger than the one of a 
single-particle property which is of the order of $2 (\tau_{c}/N t)^{1/2}$ 
\cite{allen,frenkel}. Then, long runs are needed to reach
sufficient statistics. However, the specific case of the viscosity is less 
dramatic. In supercooled liquids the viscosity is roughly proportional 
to the longest relaxation time of $F(k,t)$, 
the coherent intermediate scattering function, which occurs at $k \cong 
k_{max}$ \cite{kammerertrasl}. In particular, for 
moderately supercooled liquids a mode-coupling theory shows 
that $\eta$ is given by \cite{balucani}:

\begin{equation}
\eta = \frac{kT}{60 \pi^{2}} \int_{}^{} dt \int_{}^{} dk V^{2}(k) 
\left[\frac{F(k,t)}{S(k)}\right]^{2}
	\label{baluvisco}
\end{equation}

\noindent where $V(k) = k^{2} d \, ln \, S(k)/dk$. The vertex 
$V(k)$ greatly reduces the weight of hydrodynamic wave vectors. 
The main contributions to the above integral are due to modes being 
located around $k_{max}$ with a spread $|k_{max}-k|/\Delta \sim 1.4$,
$\Delta$ is the half-width of the main peak of $S(k)$, whose inverse 
being a measure of the extent of correlations in direct space 
\cite{balucani}. Then, for a sample of volume V$V$ the relative error 
of $\eta$ is decreased with respect to 
the above estimate by an additional factor $(v/V)^{-1/2}$, with
$v = \Delta^{-3}$ \cite{frenkel}. In order to have runs longer than 
the relaxation time of $F(k_{max},t)$ and keep reasonable 
execution times $V$ must be small. On the other hand $V$ must be larger 
than $v$. We chose samples
accommodating $N=108$ molecules and carried out runs as long as $40 
t^{\circ}$ at least, $t^{\circ}$ being the time when 
$F_{s}(k_{max},t)$ vanishes (i.e. when it drops below $0.02$). 
$t^{\circ}$ provides an estimate of the time 
scale to reach the limit in eq.\ref{einstein1}.
At the lowest temperature, $T=0.5$, the volume was 
$V \sim 150$ and $v \sim 8$. Consequently, the relative error of 
the viscosity $2 (v\tau_{c}/Vt)^{1/2}$ is 
estimated to be about $7 \%$. This figure was confirmed by 
collecting several runs at each temperature.
We also explicitely tested that the viscosity of small 
( $N=108$ ) and large samples ( $N=1000$ ) at $T=0.7$ exhibits
no significant difference . 

Small samples to evaluate $\eta$ in supercooled sustems 
were also used by Thirumalai and Mountain  \cite{thiru}.

In Fig. \ref{viscoFig} the shear viscosity is shown 
as a function of the temperature. It covers a range of more than 
three orders of magnitude. The data are also plotted as function of 
$log(T-T_{c})$  evidencing that the viscosity, differently from 
the diffusivity, may not be described by a power law analogous 
to eq.\ref{powerlaw} in the overall temperature range investigated. 
If the fit with a power law  analogous to eq.\ref{powerlaw} is 
limited to $T<0.85$, it is found $\gamma_{\eta}= -2.20 \pm 0.03$.

\subsection{The Stokes-Einstein law}

Several experimental  
\cite{rossler,sillescu1,ediger,tork2,sillescu2,tork3,voronel} 
and numerical 
\cite{antonio,ruocco,thiru,barrat,yamamoto1,sharon} works evidenced a 
decoupling of the translational diffusion and the viscosity 
on approaching the glass transition. Typically, the decoupling 
occurs around $T_{c}$ \cite{rossler}. 
To date, MD investigated the issue in one- and two-components 
{\it atomic} systems. It is therefore of interest to examine 
the present molecular system from that respect. 

The decoupling manifests as an enhancement of the 
translational diffusion $D$ with respect to the prediction of the 
Stokes-Einstein relation ( SE ) which reads \cite{lamb}

\begin{equation}
D = \frac{kT} {\eta\mu}
\label{SEEq}
\end{equation}

\noindent $\mu$ is a constant that depends on both the molecule 
geometry and the boundary conditions. For a sphere of radius $a$, 
$\mu$ equals $6\pi a$ or $4\pi a$ if stick or slip boundary 
conditions occur, respectively. The problem of uniaxial ellipsoids
in the presence of stick boundaty conditions may be worked 
out analytically \cite{lamb}. Tables of $\mu$ for prolate ellipsoids 
with slip boundary conditions were also reported \cite{tang}. 
The case of the biaxial ellipsoid 
with stick boundary condition was discussed recently 
by noting an interesting electrostatic analogy  \cite{jack1}.

\begin{figure}
\psfig{file=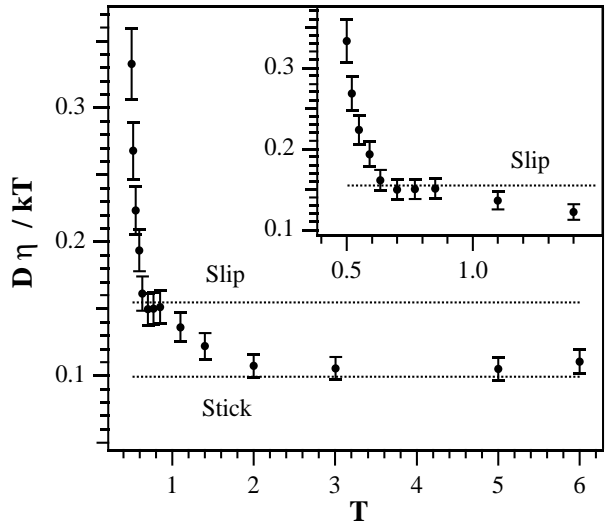,width=8cm,height=7cm}
\caption{The temperature dependence of the ratio $D\eta/kT$. 
SE predicts a constant value. Dashed lines are the SE predictions for 
prolate ellipsoids  with semiaxis $b = 0.46$ and $c =0.69$ and stick or 
slip boundary conditions. A magnification of the low-$T$ region is shown 
by the inset.}
\label{SEFig}
\end{figure}

In Fig. \ref{SEFig} we plot the ratio $D\eta / kT$ as a function
of temperature. According to SE the ratio must be constant. At higher 
temperatures the ratio levels off at about $0.105 \pm 0.007$.  
On cooling, there is first a mild change followed by a steep 
increase below $T = 0.632 = 1.38 T_{c}$.

It is believed that the SE failure is a signature of the heterogeneous 
dynamics of supercooled liquids \cite{sti,opp,douglas,sillescu3}. 
Alternative views are provided by frustrated lattice gas models 
\cite{antonio} and the ``energy landscape'' picture 
\cite{review,goldstein,sastry,angell,ruocco}. Most intepretations 
suggest the existence of crossover temperatures broadly located 
around $T_{c}$ \cite{rossler}. From this respect it is tempting 
to note that the SE law
breaks down at $T \sim 0.632$ below which intermittent behavior 
evidences ( see fig.\ref{wtdfig} ). Intermittence disappears for times
longer than $\tau_{\alpha}$ ( see fig. \ref{wtdFs} ) which is consistent 
with a growing dynamic heterogeneity of the liquid.

Around $ T=0.77$ a plateau at $0.151 \pm 0.01$ is reached. We have compared 
the results to the prediction of the SE law for prolate ellipsoids. 
The diatomic molecule under study may be roughly sketched 
as a prolate ellipsoid with semiaxis $b = 3/4$ and $c = 1/2$.
The corresponding ratio 
$D\eta / kT$ for stick and slip boundary conditions is equal to 
$0.091$ and $0.1415$, respectively. The values compare well to the 
high- and low-T plateau which are observed in fig.\ref{SEFig}. By
setting $b = 0.69$ and $c= 3/2b$ the agreement is improved 
with $D\eta / kT = 0.098$ and $0.154$ for stick and slip boundary 
conditions, respectively. 
The above analysis provides reasonable evidence of a precursors effect
of the SE breakdown which manifests itself as an apparent 
stick-slip transition. A quite similar 
crossover from stick to slip boundary conditions has been 
observed on approaching the glassy freezing of colloidal suspensions 
\cite{pusey}. Since we are studying a one-component system, 
the effect seems to be conceptually different by the apparent change 
of the boundary conditions which has been evidenced by a recent 
MD study of the motion of a guest tracer in a liquid host \cite{bagchi}.

It is worthwhile to mention that the SE law evidences fine
modifications of the supercooled liquid behavior 
which are hardly found by inspecting the diffusivity. The latter exhibits 
a single power-law regime over all the temperature range we studied.

\section{CONCLUSIONS}
\label{sec:concl}

The transport and the relaxation properties of a molecular 
supercooled liquid on the isobar $P = 1.5$ has been studied by molecular 
dynamics. The molecule is a rigid A-B system. On cooling,  
the diffusivity decrease is
fitted over four orders of magnitude by the power law
$D \propto \left( T-T_c\right)^{\gamma_{D}}$ with 
$\gamma_{D} = 1.93 \pm 0.02$ and $T_c = 0.458 \pm 0.002$. 
The divergence of the primary relaxation time $\tau_{\alpha}$ 
drawn by the intermediate scattering function $F_{s}(k_{max},t)$
tracks the diffusivity  at the lowest temperatures according to 
$\tau_{\alpha} \propto ( 
k^{2}_{max}D)^{-1}$. The result confirms findings 
on hard sphere systems \cite{fuchs}. It disagrees with other
studies on the same model system under study here which 
noticed that a power-law fit of $D$ and $\tau_{\alpha}$ yields 
the same $T_{c}$ value but $\gamma_{\alpha} > \gamma_{D}$  
with $N = 500$, $P=1$ \cite{kammerertrasl}.

At the lowest temperatures fractions of highly mobile and trapped 
molecules are evidenced, then extending previous results on supercooled 
atomic mixtures to one-component molecular liquids \cite{thiru,claudio}.
Translational jumps are evidenced. The duration 
of the jumps exhibits a distribution. The distribution of the 
waiting-times before a new jump takes place, $\psi(t)$, is exponential 
at higher temperatures. At lower temperatures two regimes are 
evidenced: at short times $\psi(t) 
\propto t^{\xi-1}$ with $ 0 < \xi \leq 1$ whereas at long times 
the decay is faster than exponential. The crossover between the two 
regimes occurs around $\tau_{\alpha}$. Noticeably in this time window 
MCT predicts that $F_{s}(k_{max},t)$ also decays as a power law. The 
interplay between the two time fractals is anticipated and a 
preliminary analysis has been attempted.
The fractal distribution of the waiting time is ascribed to 
the intermittent behavior which is expected to develop in glassy 
systems \cite{sjogren,hubbard,odagaki,sharon}. If the waiting time 
before a new jump exceeds $\tau_{\alpha}$ the environment surrounding 
each molecule largely restructures. The subsequent average process
results in a faster decay of $\psi(t)$. 
The probability that no jump occurred before time $t$ is found to 
vanish  slower than $F_{s}(k_{max},t)$ in close analogy with the 
case of viscous silica melts \cite{silica}. 

The shear viscosity has been studied over more than three orders of 
magnitude. The data are fitted by the power law 
$\eta \propto \left( T-T_c\right)^{\gamma_{\eta}}$ with 
$\gamma_{\eta}= -2.20 \pm 0.03$ at the lowest temperatures. The 
validity of the Stokes-Einstein relation has been examined. Previous 
MD studies were limited to atomic one- and two- components systems.
At higher temperatures SE 
fits well the data if {\it stick} boundary conditions are assumed. 
At lower temperatures the product $D\eta/T$ increases and 
the Stokes-Einstein relation is not obeyed. The breakdown occurs near 
to the temperature where the intermittency is evidenced by $\psi(t)$.
Interestingly, a precursor effect of the breakdown is observed which 
manifests as an apparent stick-slip transition. A 
crossover from stick to slip boundary conditions has been 
observed on approaching the glassy freezing of colloidal suspensions 
\cite{pusey}.

\acknowledgments

The authors warmly thank Walter Kob for having suggested the 
investigation of the present model system, the careful reading of 
the manuscript and the comments on it. Umberto Balucani, Claudio Donati 
and Francesco Sciortino are thanked for many helpful discussions and
Jack Douglas for a preprint of ref. \cite{sharon}.

\appendix
\section{RATTLE-NPT algorithm}
\label{app:NPT}

The well-known Nos\`e-Andersen algorithm ensures NPT equilibrium, 
i.e. equilibrium  at constant number of particles, pressure $P$ and 
temperature $T$ \cite{allen}. It defines one extra degree of freedom 
$s(t)$, describing the thermal bath and allows the change of the 
volume $V(t)$.  Since the algorithm was originally derived for 
atoms, the extension to molecules is of interest. To this aim, 
the so-called constraint methods are of help, particularly the 
RATTLE algorithm \cite{allen}. It evaluates the 
dynamics of polyatomic molecules by defining proper forces  
constraining the relative motion of the atoms belonging to the same 
molecule. Below, the combined RATTLE-NPT algorithm is described.
 
First, the vector ${\bf R}(t+\dt)$ of all 
atoms positions, $s(t+\dt)$ and the vector $V(t+\dt)$ of 
all atoms velocities and the their first time  derivatives  at mid-step 
${\bf V}(t + \dt /2$,${\dot s}(t+\dt/2)$ , ${\dot V}(t+\dt/2)$ 
are evaluated  according to the RATTLE algorithm . 
The forces  ${\bf f}_{i}(t+\dt)$ (i refers to the i-th atom) 
are expressed in terms of the configuration ${\bf R}(t+\dt)$.  
The other steps are the following:

\begin{enumerate}
\item Guess of the first derivatives of interest at time $t+\dt$, 
$\dot y^g(t+\dt)$ ( $\dot y = {\bf v}, \dot s, \dot V$) :

$$
\dot y^g(t+\dt) = 2 \dot y(t) - \dot y(t-\dt) 
$$

The above guess, which is correct to second order, is not unique and 
alternatives are possible.
\item Calculation of temperature $T(t+\dt)$ by  using ${\bf v}^g(t)$.
\item Calculation of  $\dot s(t+\dt)$. Replacing the lagrangian 
equation for $\ddot s(t+\dt)$ into the velocity Verlet equations 
\cite{allen} yields:

$$
\dot s(t+\dt) = A + \frac{A^2}{s} \dt + \frac{A^3}{2s^2} \dt^2 + 
O(\dt^3),
$$

with $A = s(t+\frot\dt) + \frot\dt f \frac{T(t + \dt) - T_{0}}{Q} 
s(t+\dt)$. $Q$, $T_0$ and $f$ are the mass of the thermal piston, the 
thermal bath temperature and the degrees of freedom 
( $f = \frac{5}{2} N_{at} - 3$ in the present case ), respectively. 

\item Calculation of pressure $P(t+\dt)$.

\item Evaluation of ${\dot V}(t+\dt)$ according to the velocity 
Verlet algorithm:

\begin{eqnarray*}
\dot V&&(t+\dt) = \\
&&\frac{\dot V(t+\frot\dt) + \frot\dt 
 \frac{s^2(t+\dt)}{W}(P(t+\dt) - P_{0})}{1 - \frot\dt \dot 
s(t+\dt)/s(t+\dt)}
\end{eqnarray*}

where $W$ and $P(t)$ are the mass of the piston setting the pressure 
at $P_0$ and the instantaneous pressure, respectively.

\item 
Evaluation of the velocities at time $t+\dt$, ${\bf v}_i(t+ 
\dt)$ in the presence of the intermolecular forces, the thermal 
bath forces and the forces due to the mechanical piston.
According to the velocity Verlet algorithm:

\begin{eqnarray*}
&&{\bf v}_i(t + \dt) =  (1+ \frot\dt\dot s(t+\dt)/s(t+\dt))^{-1} 
\times \\
&&\Bigg \{ {\bf v}_i(t+\frot\dt) + \frot \dt {\bf f}_i(t +\dt) +\\
&&\frot \dt 
 \left[ \frac{\ddot V}{3V} + \left ( \frac{\dot s}{s} 
-\frac{2}{3}\frac{\dot V}{V} \right )
 \frac{\dot V}{3 V}\right] {\bf R}_{cm,i} \Bigg \}
\end{eqnarray*}

where ${\bf R}_{cm,i}$ is the center of mass of the molecule where  
the i-th atom is located.

\item Adjustement of the velocity ${\bf v}_i$. The relative 
velocities of atoms which are mutually bonded are considered and 
their component along the bond is made to vanish, according to the 
RATTLE algorithm.

\end{enumerate}

Finally we note that iterating steps from $2$ through $7$ increases 
the accuracy considerably. The present algorithm  needs to store 
the arrays  ${\bf V}(t-\dt), \dot s(t -\dt)$ and $V(t-\dt)$ to  
guess the velocity as $v^g(t+\dt)$ in the first step.


\begin{references}
\bibitem[*]{byline}  corresponding author: e-mail address: 
leporini@mailbox.difi.unipi.it.
\bibitem{noi} Proceedings of the II Workshop on Non-Equilibrium Phenomena 
in Supercooled Fluids, Glasses and Amorphous Materials, M.Giordano,
D.Leporini, M.P.Tosi ( edts.). {\it  J.Phys.:Condens.Matter}, 
Vol.11, No.10A (1999).
\bibitem{review} for a short review see: 
M.D.Ediger, C.A.Angell, S.R.Nagel {\it J.Phys.Chem} {\bf 100}, 13200 (1996);
\bibitem{rossler} E. R\"{o}ssler, Phys.Rev.Lett. \bf 65 \rm, 1595 (1990).
\bibitem{sillescu1} F. Fujara, B.Geil, H.Sillescu, G.Fleischer
Z.Phys. {\bf B88}, 195 (1992); I.Chang, F.Fujara, B.Geil, G.Heuberger, 
T.Mangel, H.Sillescu J.Non-Cryst.Solids {\bf 172-174} 248 (1994).
\bibitem{rosslerjpc} E. R\"{o}ssler, J.Tauchert, P.Eiermann
J.Phys.Chem. \bf 98 \rm, 8173 (1994).
\bibitem{rosslerjcp} E. R\"{o}ssler, P.Eiermann J.Chem.Phys. {\bf 100}, 
5237 (1994).
\bibitem{ediger} M.T. Cicerone, F.R.Blackburn, M.D.Ediger{\it J.Chem.Phys.},
{\bf 102}, 471 (1995); M.T. Cicerone, M.D.Ediger {\it J.Chem.Phys.},
{\bf 104}, 7210 (1996).
\bibitem{tork2} J.C.Hooker, J.M.Torkelson {\it Macromolecules}, 
{\bf 28},7683 (1995).
\bibitem{sillescu2} G.Heuberger, H.Sillescu {\it J.Phys.Chem.}, 
{\bf 100}, 15255 (1996).
\bibitem{ye} J.Y.Ye, T.Hattori, H.Nakatsuka, Y.Maruyama, M.Ishikawa, 
{ \it Phys.Rev.B} {\bf 56}, 5286 (1997).
\bibitem{tork3} D.B.Hall, A.Dhinojwala, J.M.Torkelson {\it Phys.Rev.Lett.}
 {\bf 79}, 103 (1997).
\bibitem{lepo2} L.Andreozzi, A.Di Schino, M.Giordano, D.Leporini,  
{ \it Europhys.Lett.}  {\bf 38}, 669  (1997).
\bibitem{voronel} A.Voronel, E.Veliyulin, V.Sh.Machavariani,
A.Kisliuk, D.Quitmann  {\it Phys.Rev.Lett.} {\bf 80}, 2630 (1998).
\bibitem{lepo3} M.Faetti, M.Giordano, L.Pardi, D.Leporini 
Macromolecules, {\bf 32}, 1876 (1999).
\bibitem{lepo4} L.Andreozzi, M.Faetti, M.Giordano, D.Leporini 
J.Phys.Chem.B {\bf 103} 4097 (1999).
\bibitem{sti} J. A. Hodgdon, F. H. Stillinger Phys.Rev.E {\bf 48}, 
207 (1993); F. H. Stillinger, J. A. Hodgdon ibid. {\bf 50}, 2064 (1994); 
\bibitem{opp}C.Z.-W. Liu, I.Oppenheim,{\it Phys.Rev.} {\bf E53}, 799 (1996).
\bibitem{douglas} J.F.Douglas, D.Leporini {\it J.Non-Cryst.Solids} 
{\bf 235-237}, 137 (1998).
\bibitem{sillescu3} H.Sillescu {\it J.Non-Cryst.Solids} {\bf 243}, 
81 (1999).
\bibitem{antonio} M. Nicodemi, A. Coniglio, {\it Phys. Rev. E }{\bf 
56}, R39 (1998); A.Coniglio, A.DeCandia, A.Fierro, M. Nicodemi
{\it  J.Phys.:Condens.Matter}, {\bf 11}, A167 (1999).
\bibitem{goldstein} M.Goldstein {\it J. Chem. Phys.} {\bf 51}, 3728 
(1969); 
\bibitem{sastry} S.Sastry, P.G.Debenedetti, F.H.Stillinger {\it Nature} 
{\bf 393}, 554 (1998); 
\bibitem{angell} C.A.Angell {\it J.Res.Natl.Inst.Std.Tech.},{\bf 102}, 171 
(1997); C.A.Angell, B.E.Richards, V.Velikov {\it  J.Phys.:Condens.Matter}, 
{\bf 11}, A75 (1999).
\bibitem{ruocco} L.Angelani,G.Parisi,G.Ruocco,G.Viliani 
{\it Phys.Rev.Lett.} {\bf 81}, 4648 (1998).
\bibitem{gotze} W.G\"{o}tze {\it  J.Phys.:Condens.Matter}, 
{\bf 11}, A1 (1999); H.Z.Cummins {\it  J.Phys.:Condens.Matter}, 
{\bf 11}, A95 (1999).
\bibitem{gotze2} T.Franosch, W.G\"{o}tze {\it Phys. Rev. E }{\bf 57}, 
5833 (1998).
\bibitem{kob} W.Kob, {\it  J.Phys.:Condens.Matter} {\bf 11} R85 (1999).
\bibitem{thiru} D.Thirumalai, R.D.Mountain {\it  Phys. Rev. E }{\bf 
47}, 479 (1993).
\bibitem{barrat} J.-L.Barrat, J.-N.Roux, J.P.Hansen {\it Chem.Phys.}, 
{\bf 149}, 197 (1990).
\bibitem{yamamoto1} R.Yamamoto, A.Onuki {\it Phys. Rev. E }{\bf 
58}, 3515 (1998).
\bibitem{yamamoto2} R.Yamamoto, A.Onuki {\it Phys. Rev. Lett. }{\bf 
81}, 4915 (1998).
\bibitem{harrowell} D.Perera and P.Harrowell {\it Phys. Rev. Lett. }{\bf 
81}, 120 (1998).
\bibitem{sharon} P.Allegrini, J.F.Douglas, S.H.Glotzer submitted to 
Phys.Rev.E
\bibitem{claudio} W.Kob, C.Donati, S.J.Plimpton, P.H.Poole, S.C.Glotzer
Phys.Rev.Lett. {\bf 79} 2827 (1997); C.Donati, j.F.Douglas, W.Kob, 
S.J.Plimpton, P.H.Poole, S.C.Glotzer Phys.Rev.Lett. {\bf 80} 2338 (1998).
\bibitem{bernu} H.Miyagawa, Y.Hiwatari, B.Bernu, J.P.Hansen, 
 J.Chem.Phys. {\bf 88},3878 (1988);
\bibitem{wahnstrom} G.Wahnstr\"{o}m, Phys.Rev.A, {\bf 44}, 3752 (1991).
\bibitem{muranaka} T.Muranaka, Y.Hiwatari, J.Phys.Soc of Japan, 
{\bf 67} 1982 (1998).
 \bibitem{demichel2} C.De Michele, D.Leporini submitted to Phys. Rev. E.
\bibitem{kammererrot1} S.K\"{a}mmerer, W.Kob, R.Schilling Phys.Rev.E 
{\bf 56} 5450 (1997).
\bibitem{kammererrot2} S.K\"{a}mmerer, W.Kob, R.Schilling Phys.Rev.E 
{\bf 58} 2141 (1998).
\bibitem{kammerertrasl} S.K\"{a}mmerer, W.Kob, R.Schilling Phys.Rev.E 
{\bf 58} 2131 (1998).
\bibitem{allen} M.P.Allen, D.J.Tildesley {\it Computer Simulation of 
Liquids} (Clarendon Press, Oxford 1987 )
\bibitem{private} S.K\"{a}mmerer, W.Kob, R.Schilling, private communication
\bibitem{hansen}J.-P.Hansen, I.R. McDonald {\it Theory of Simple 
Liquids} II Edition (Academic Press, London 1986).
\bibitem{fuchs} M.Fuchs, I.Hofacher, A.Latz  Phys.Rev A{\bf 45} 898 (1992).
\bibitem{harrowell2} M.M.Hurley, P.Harrowell J.Chem.Phys.{\bf 105} 
10521 (1996).
\bibitem{sjogren} L.Sj\"{o}gren {\it Z.Phys.B} {\bf 74}, 353 (1989).
\bibitem{hubbard} J.F.Douglas, J.B.Hubbard {\it Macromolecules} 
{\bf 24}, 3163 (1991); J.F.Douglas Comp.Mat.Sci. {\bf 4}, 292 (1995).
\bibitem{odagaki} T.Odagaki Phys.Rev.Lett. {\bf 75}, 3701 (1995).
\bibitem{hilfer} R.Hilfer, L.Anton {\it Phys.Rev.E} {\bf 51}, R848 (1995).
\bibitem{silica} J.Horbach, W.Kob {\it Phys.Rev.B} {\bf 60}, 3169 (1999).
\bibitem{allen1} M.P.Allen, D.Brown, A.J.Masters Phys.Rev.E {\bf 49} 2488 
(1994).
\bibitem{note} In principle, the viscosity $\eta$ may be also evaluated 
by a proper Green-Kubo relation involving the area below the 
self-correlation function of the off-diagonal components of the 
pressure tensor ${\cal P}_{\alpha \beta}$, $\eta(t)$  \cite{hansen,allen1}. 
However, the slowly-decaying tail of $\eta(t)$  poses numerical 
problems \cite{barrat,allen,hansen}.
\bibitem{allen2} M.P.Allen Mol.Phys. {\bf 52} 705 (1984).
\bibitem{cui} S.T.Cui, P.T.Cummings, H.D.Cochran Mol.Phys. {\bf 88} 1657 
(1996).
\bibitem{frenkel} D.Frenkel in Proceedings of the International School of 
Physics "Enrico Fermi", Vol. 75 ( Soc. Italiana di Fisica, Bologna 1980 ).
\bibitem{balucani} U.Balucani, Mol.Phys. {\bf 71} 123 (1990); 
S.Bhattacharyya, B.Bagchi J.Chem.Phys. {\bf 109},7885 (1998).
\bibitem{lamb} H.Lamb {\it Hydrodynamics} VI Ed. ( Cambridge University 
Press, Cambridge 1932 ).
\bibitem{tang} S.Tang, G.T.Evans Mol.Phys. {\bf 80} 1443 (1993).
\bibitem{jack1} J.B.Hubbard, J.F.Douglas, {\it Phys.Rev.E} {\bf 47}, 
R2983 (1993).
\bibitem{pusey} P.N.Segre', S.P.Meeker, P.N.Pusey, W.C.K.Poon 
Phys.Rev.Lett. {\bf 75}, 958 (1995).
\bibitem{bagchi} G.Srinivas, S.Bhattacharyya, B.Bagchi J.Chem.Phys. {\bf 
110}, 4477 (1999).
\end{references}
\end{document}